\documentclass[final
  ]
  {aipproc}

\layoutstyle{6x9}

\usepackage{graphics,latexsym,amsmath,epsfig,bm}

\newcommand{\as}{\alpha_s}
\newcommand{\e}{\mathrm{e}}
\newcommand{\shat}{\hat{s}}
\renewcommand{\d}{\mathrm{d}}
\newcommand{\A}{\mathcal{A}}
\newcommand{\B}{\mathcal{B}}
\newcommand{\C}{\mathbf{C}}

\newcommand{\Sa}{\A_{1,S}(0)}
\newcommand{\FO}{\A_{8}(0)}
\newcommand{\FOQ}{\A_{8}(Q_0)}
\newcommand{\FSQ}{\A_{1}(Q_0)}
\newcommand{\FS}{\A_{1}(0)}
\newcommand{\mbQ}{\B(Q_0)}
\newcommand{\FL}{\A_{BFKL}}
\renewcommand{\Im}{\mbox{$\mathrm{I\!Im}$}}

\begin{document}

\title{Gaps between Jets: Matching two Approaches
}


\author{A. Kyrieleis}{
 address={University of Manchester, 
  Oxford Road, Manchester M13 9PL, U.K.\\
 email: kyrieleis@hep.man.ac.uk}
}

\begin{abstract}
We calculate the parton level cross section for the 
production of two jets that are far apart in rapidity, subject to a limitation 
on the total transverse momentum $Q_0$ in the interjet region. 
We specifically address the question of how to combine the approach which
sums all leading logarithms in $Q/Q_0$ (where $Q$ is the jet transverse momentum)
with the BFKL approach, in which leading logarithms of the scattering energy are
summed.  Using an ``all orders'' matching, 
we obtain results for the cross section which correctly reproduce the two
approaches in  the appropriate limits. 
\end{abstract}

\maketitle


\section{Introduction}
Final states with high-$p_T$ jets separated by large rapidity gaps at
hadron colliders offer the possibility to better understand QCD in
the high energy limit and also to understand QCD radiation in ``gap'' events.
There are two major approaches to the production of two gap-separated
jets.
In the BFKL \cite{BFKL} approach, parton-parton elastic scattering with a QCD
colour singlet exchange is  regarded as providing
the leading contribution to the cross-section.
The  leading-$Y$ terms ($Y$ is the rapidity
interval between the jets) are summed  i.e. terms  $ \sim \as^n Y^n$ \cite{BFKL
  app, MoMaRy}.  The observable calculated in this approach does not consider any
radiation into the interjet region. Experiments though, impose an
upper bound  on this radiation by necessity.
In the second approach  soft radiation with
transverse energy below $Q_0$ is allowed
in the interjet region. This gives rise to  logarithms of $Q/Q_0$ where
$Q$ is the transverse momentum of the jets.
The global leading  logarithms of $Q/Q_0$ ($LLQ_0$) have  been summed for various jet 
definitions \cite{sterman,AbSey03} i.e terms $\sim \as^n Y^m
L^n\;  (m\le n) $  where  $L=\ln Q^2/Q_0^2$ . Non-global effects have been
considered in \cite{AbSey03,AbSey02}.
In order to get a better understanding  of
the gaps-between-jets processes  at colliders it is desirable to combine the
two approaches. This is the  main issue in this contribution, for details see
\cite{ours} 
\vspace{-.25cm}
\section{Summing logarithms in $Q_0$}
As the first step we  recalculate the cross section for two-jet production in the
high-energy (i.e.~high rapidity separation) limit, with limited total
scalar transverse momentum in the interjet region. We require this
transverse momentum to be below $Q_0$ and consider
the region $ Q_0^2 \ll Q^2 \ll \shat=\e^Y\,Q^2$.
 Since we are not
sensitive to collinear emission, we work at the parton level and calculate
the all-orders  gap cross section $\sigma\equiv\frac{\d\sigma(\shat,Q_0,Y)}{\d Q^2}$
for the process  $\mathrm{qq'\to qq'}$. $\sigma^{(n)}$ denotes the
cross section at $\mathcal{O}(\as^n)$.  Our approximation implies the eikonal
 (soft gluon) approximation. 
To generate the leading logs in $Q_0$, we make the approximation
of strongly-ordered  transverse momenta for real and virtual gluons. 

As the basis for the calculation to all
orders we employ the following theorem.
Let us  denote by  $\A_1^{(n)}(Q_0) \C_1 \:(\A_8^{(n)}(Q_0) \C_8) $  the singlet
(octet) component of the $\mathcal{O}(\as^n)$  $\mathrm{qq'\to qq'}$ amplitude
in the  approximation defined above, with the phase space for the gluons
constrained to the gap region in rapidity and with transverse momentum above~$Q_0$  ($\C_{1,8}$ are the  colour factors).  With
$\B(Q_0)$ denoting the production amplitude (including colour factor) for
more than 2 particles, the theorem reads
\begin{align}
\sigma^{(k)} = |\FS|^2\C_1^2 + |\FO|^2\C_8^2 + |\mbQ|^2 =
|\FSQ|^2\C_1^2 + |\FOQ|^2\C_8^2  
\label{theorem}
\end{align}
where the squares are to be read symbolically representing the sums over
$\A^{(n)*}\A^{(m)}$ (and $\B(Q_0)$ , respectively).
This is clearly a major simplification, since it means that we never have to
calculate any real emission or triple-gluon-vertex diagrams. This theorem 
provides the basis for the matching with BFKL.
We calculate  $\A(Q_0)_{1,8}$ and hence $\sigma$  to all orders. 
Besides the 
double-leading-logarithmic ($DLL$) terms we include
those terms sub-leading in $Y$ that arise from the imaginary parts of the
loop integrals. 
\vspace{-.2cm}
\section{Matching with BFKL}
To combine the gap cross section with the BFKL approach order-by-order we need to
prevent double counting and make sure the divergences arising from the BFKL
approach (at each order in $\alpha_s$) cancel in the jet cross section. To
this end we calculate the leading-$Y$ approximation of the singlet component
$\A_1^{(n)}(Q_0)$: 
\begin{align}
 \A^{(n)}_{1,S}(Q_0) \equiv \left.\A^{(n)}_1(Q_0)\right|_{LY}.
\end{align}
$\A^{(n)}_{1,S}(0)$ is divergent at each order and it is this contribution to
$\sigma$ that is also included in the BFKL result.  
\vspace{-.1cm}
\paragraph{Fixed order matching}
We denote by $\A^{(n)}_{BFKL} \C_1$ the $\mathcal{O}(\as^n)$ elastic quark
scattering amplitude with colour singlet exchange in the leading-Y
approximation.  We want  $\sigma$
 to include $\A^{(n)}_{BFKL}$. However, $\A^{(n)}_1(0)$ also includes terms
sub-leading in $Y$  which we have to keep; they are given by
$(\A_1(0)-\A_{1,S}(0))^{(n)}$. We therefore define the following fixed order gap
cross section (again omitting the sum over indices in the first line).
\begin{align}
\sigma^{(k)}_{gap} &\equiv |\FL + \FS - \Sa |^2 \C_1^2+
|\FO|^2\C_8^2 + |\mbQ|^2\\ 
&=\sigma^{(k)} +  \sum_{m+n=k} \left[ 2
\Im\A^{(m)}_1(0)\cdot(-i \delta^{(n)}) + \delta^{(m)}\delta^{(n)*}\right]
\C_1^2  \label{fix com}\\
&\mbox{with }\quad \delta^{(n)}=\FL^{(n)}-\A^{(n)}_{1,S}(0)\label{delta}
\end{align}
where, in the last  line we have invoked the theorem
\eqref{theorem}. This cross section combines the two approaches without
double counting. However, not surprisingly, the strong ordering approximation
cannot cancel the divergence in the BFKL amplitude at any order. The second
term in \eqref{fix com} and hence  $\sigma^{(k)}_{gap} $ is
divergent for $k\ge 6$. Via \eqref{fix com} we can therefore combine the
all-orders cross section $\sigma$  with the BFKL result up to $\mathcal{O}(\as^5)$.

The  theorem \eqref{theorem} holds beyond the high energy
approximation, the matching with BFKL can therefore be extended to full (global) $LLQ_0$ accuracy in a straightforward way \cite{ours}. 
\vspace{-.1cm}
\paragraph{All orders matching}
Although the  order-by-order combination of the  $LLQ_0$ and the BFKL result
can only work for the first few orders it is possible to construct an
all-orders cross section that does smoothly interpolate the $LLQ_0$ and BFKL
results, agreeing with each in its region of validity and avoiding any
double-counting. Central to this are the following two
observations. First, the amplitude $\A^{(n)}_{1,S}(Q_0)$ summed to all orders
reads: 
\begin{align}
\A_{1,S}(Q_0)= -i\,\frac{N_c^2-1}{2N_c^3}\;\frac{\pi}{Y} \; \A^{(1)}_8
\cdot\left[ 1 - \exp\left(-\frac{N_c \as}{2 \pi} \;Y L\right) \right].
\label{singl resum}
\end{align} 
The exponential vanishes as $Q_0\to 0$.  In contrast to the fixed order result,
$\A_{1,S}(0)$ is therefore finite. Secondly, we find the following relation
between the (finite) all-orders results  for
the BFKL $2\to 2$ cross section $\sigma_{BFKL}$ \cite{MoMaRy} and the  gap
cross section $\sigma$: 
\begin{align}
\left.\sigma_{BFKL}\right|_{Y\to 0}  = \left.\sigma\right|_{Y\to \infty} =
\sigma_{S} \equiv |\A_{1,S}(0)|^2 \C_1^2 =
\sigma^{(2)}\;\frac{N_c^2-1}{N_c^4}\; \frac{\pi^2}{Y^2} \label{fund rel}
\end{align}
which implies $\A_{BFKL}|_{Y\to 0}  = \A_{1,S}(0)$. Using these two remarkable
results we construct three different matched cross sections ($\delta$ is given by
\eqref{delta} summed to all orders).
\begin{description}
\item[Simple matching: $\quad \sigma_{gap} = \sigma + N_c^2 |\delta|^2$]
\item[Cross section matching:$\quad  \sigma_{gap} =\sigma + \sigma_{BFKL} -
  \sigma_{S}$ ]
\item[Amplitude matching: $\quad  \sigma_{gap} = \frac14(N_c^2-1)|\A_8(Q_0)|^2 +
  N_c^2|\A_1(Q_0)+\delta|^2$]
\end{description}
 In the first scheme we have replaced all expressions in  \eqref{fix com} with  the
(finite)  all-orders results and exploited the fact that $\A_1(0)$ is zero. In
all three cases we subtract from the sum of the $LLQ_0$ and $BFKL$ amplitudes 
(cross sections) the double-counted term $\A_{1,S}(0)$ ($\sigma_s$). In
all schemes $\sigma_{gap}\to \sigma$ for  $Y\to 0$ since $\delta\to 0\; 
(\sigma_{BFKL}-\sigma_S\to 0)$, see \eqref{fund rel}.  As  $Y\to \infty$
  we have $\sigma, \sigma_S \to 0$ and $\A_{1,8}(Q_0), \A_{1,S}(0)\to 0$
  (i.e. $\delta\to \A_{BFKL}$) 
 and  hence  $\sigma_{gap}\to \sigma_{BFKL}$.
 Each scheme
therefore achieves our goal of having a smooth 
matching of the two all-orders cross sections, in that for small and
large $Y$  it agrees with the $LLQ_0$ and BFKL cross sections
respectively avoiding any double-counting.

\begin{figure}
  \includegraphics[height=.3\textheight]{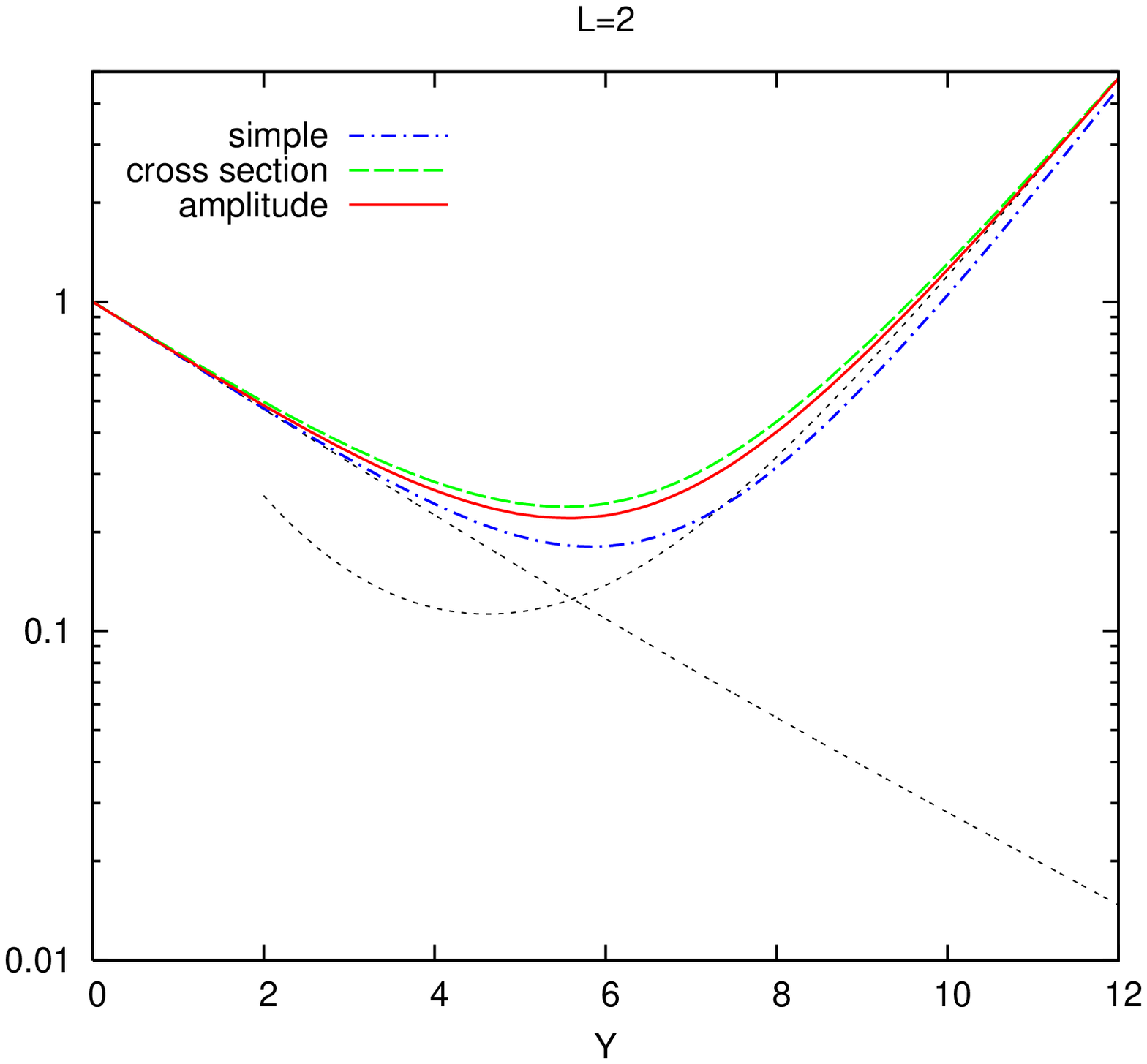}
  \includegraphics[height=.3\textheight]{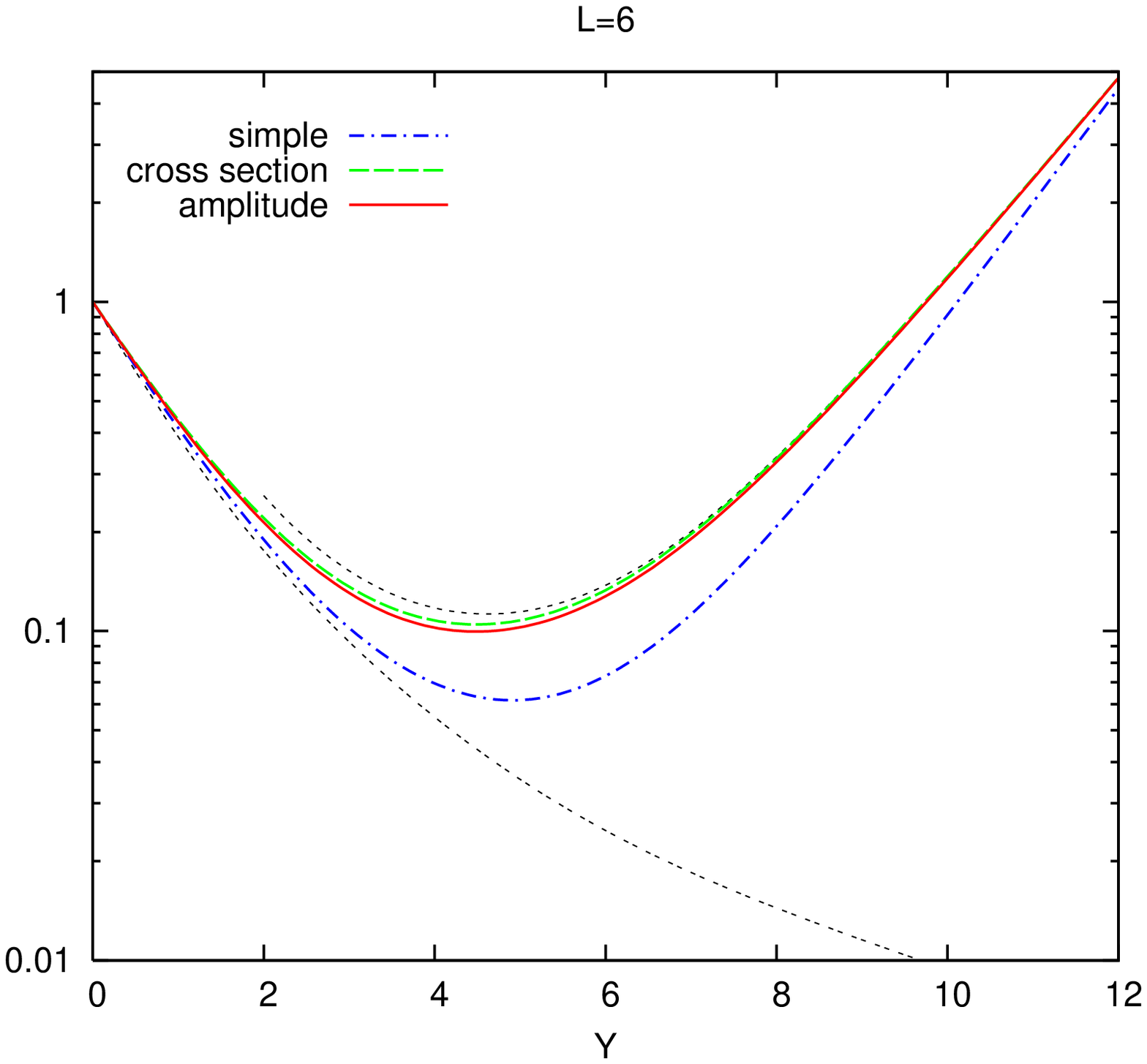}
\caption{The gap cross section in the
three matching schemes for $L=2$ and $6$ ($\as=0.2$) compared to
$\sigma_{BFKL}$ (dots) and $\sigma$ (double-dots)\label{comb}}
\end{figure}

 As a measure of the uncertainty inherent in the matching procedure
 fig. \ref{comb} shows numerical results of all three schemes. Indeed, they
 all match the two cross sections in the small and large $Y$ limits and the
 differences are not large in between.  
\vspace{-.2cm}
\section{Conclusion}
Working in the high energy limit we have calculated the (partonic)
cross section for the production of two jets distant in rapidity 
and with limited transverse energy flow into the  region between the
jets. Besides the $DLL$  terms, we have summed
terms sub-leading in $Y$ stemming from the imaginary parts of the loop
integrals. This allowed us to consistently combine
the terms of the $LLQ_0$ series and the BFKL series to
$\mathcal{O}(\as^5)$ accuracy without double counting. In the 
$LLQ_0A$, the inclusion of higher orders of the BFKL cross section
in this way is not possible since it implies a divergent cross
section. 

We have also studied several ``all order'' matching schemes that
effectively interpolate between the $LLQ_0$ and BFKL results.  Although
they all yield similar results, the differences between them cannot be
resolved without further work, specifically understanding the role of
real-emission contributions in the high energy limit.
We have made a first step towards the unification of the two
main approaches to the ``jet--gap--jet'' process.
\vspace{-.3cm}
\begin{theacknowledgments}
The presented work was done in collaboration with J. R. Forshaw and
M.~H.~Seymour. 
\end{theacknowledgments}

\end{document}